\documentclass[review]{elsarticle}

\usepackage{verbatim}
\usepackage{physics}
\usepackage[normalem]{ulem}
\usepackage[dvipsnames]{xcolor}
\usepackage{siunitx}
\usepackage{lineno,hyperref}
\usepackage[scale=0.7,top=1cm,bottom=3cm]{geometry}

\begin{document}

\begin{frontmatter}

\title{Chaotic Behavior of Quantum Cascade Lasers at Ignition}

\author[1]{D. E. \"{O}nder \corref{cor1}}
\ead{ekin.onder@teorfys.lu.se}
\author[1]{A. A. S. Kalaee}
\author[2]{D. O. Winge}
\author[1]{A. Wacker}
\ead{Andreas.Wacker@fysik.lu.se}

\affiliation[1]{organization={Mathematical Physics and Nanolund, Lund University},
addressline={Box 118},
postcode={22100},
city={Lund},
country={Sweden}}
\affiliation[2]{organization={Synchrotron Radiation Physics and Nanolund, Lund
    University},
addressline={Box 118},
postcode={22100},
city={Lund},
country={Sweden}}

\cortext[cor1]{Corresponding author}

\begin{abstract}
The ignition of Quantum Cascade Lasers can occur from a state of oscillating field domains. Here, the interplay between lasing and the kinetics of traveling domain boundaries provides complex oscillation scenarios. We analyze our numerical findings in detail for a device operating at terahertz frequencies and manifest chaotic evolution by positive Lyapunov exponents. This shows that these important devices can exhibit chaotic behavior even without periodic driving, which needs to be taken into account in their design. 
\end{abstract}

\begin{keyword}
  Quantum Cascade Laser \sep Chaos \sep Simulation \sep Lyapunov exponents
\end{keyword}

\end{frontmatter}



\section{\label{sec:Intro}Introduction}
 Negative differential conductivity (NDC), i.e., the decrease of current with increasing electric field, is a common source of instabilities in semiconductor devices \cite{ShawBook1992,SchollBook2001,BonillaBook2010}. In extended systems like the Gunn diode\cite{GunnIBM1964}, it leads to the formation of spatial domains  with different electric fields. Commonly, these travel through the device causing characteristic oscillations
\cite{KroemerIEEETransElectronDev1966}. A related system are semiconductor superlattices\cite{EsakiIBM1970,WackerPhysRep2002,BonillaRepProgPhys2005}, where a wide scenario of stationary, oscillating, and chaotic evolution\cite{LuoPRL1998,AmannPRL2003} was studied.

\begin{figure}
\centering
\includegraphics[scale=0.4]{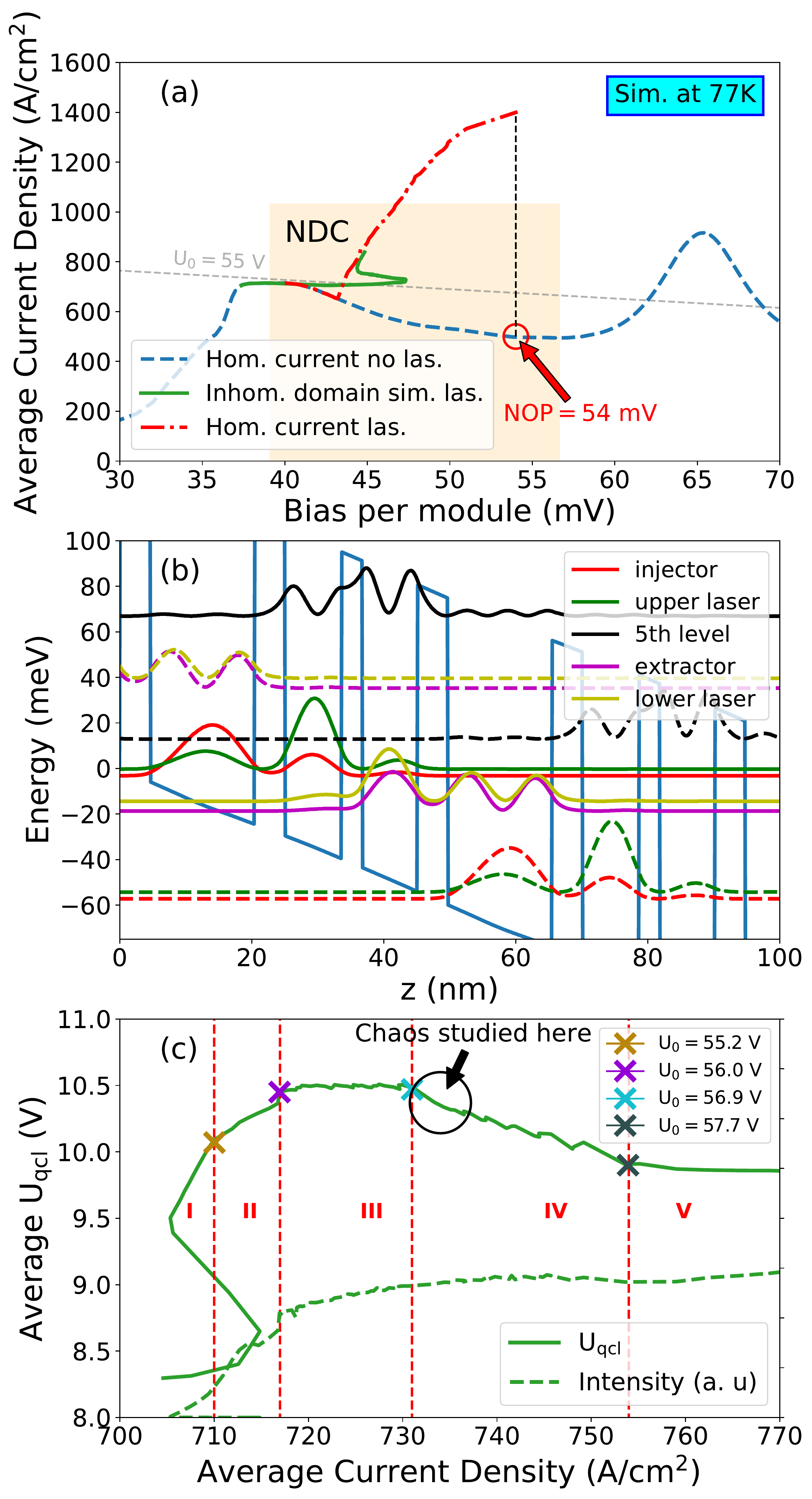}
\caption{(a) Simulated current bias relation with and without irradiation. The broken curves assume a homogeneous bias drop over the structure. Without lasing (blue dashed line) the NOP is located in a NDC region. Taking into account the self-consistent lasing field (red dash-dotted line), the current is strongly enhanced and stabilizes the NOP. The full green curve shows the result for domain formation including the self-consistent lasing field. Here the device is driven by an external bias $U_0$, which defines the load line (dashed gray) given by Eq.~\eqref{eq:current}.
(b) Blue line: Heterostructure potential for a layer sequence \textbf{46} /158/ \textbf{46} /86/ \textbf{31.7}/83 $\si{\angstrom}$ with $\mathrm{Al_{0.15}Ga_{0.85}As}$  barriers (in boldface) and GaAs wells for the device V812 from \citep{FathololoumiJAP2013}.
The Wannier-Stark states at the NOP are shown by different colors, where full lines denote states in the central module $0<z\le d=45.07\,\textrm{nm}$, and dashed lines are attributed to neighbouring modules.
(c) Light bias current density (LUJ) characteristics under domain formation. Red vertical lines represent boundaries between regions of different behavior occurring for values of $U_0$ specified in the upper right inset.
}
\label{fig:average}
\end{figure}

The Quantum Cascade Laser (QCL)\cite{FaistScience1994,FaistBook2013} is currently the most important device for mid and far-infrared radiation. QCLs are based on carefully designed semiconductor heterostructures, see, e.g., Fig.~\ref{fig:average}(b). These guide the electron flow by tunneling and scattering
to establish electronic inversion for a pair of quantum levels (the laser levels) at a
specific electric field, the nominal operating point (NOP). In order to increase the total gain, a module of several layers including the laser levels is repeated several times, so that the electrons traverse the total structure like water in a cascade. Thus, the field distribution in QCLs exhibits domain formation if driven in an NDC region \cite{LuPRB2006,DharSciRep2014,AlmqvistEurPhysJB2019}.
As the devices are most efficient, if all modules contribute equally to the gain, it is a common strategy to avoid NDC around the NOP in the QCL-design.
On the other hand, resonant tunneling is prone to provide NDC above alignment \cite{KazarinovSPS1971,EsakiPRL1974,CapassoAPL1986}. Therefore, NDC is ubiquitous in layered structures such as QCLs and instabilities close to threshold \cite{FathololoumiJAP2013,KhabibullinOptoElectronicsReview2019} are not always avoidable.

Here we focus on device V812 from \citep{FathololoumiJAP2013}, a QCL operating with good performance at terahertz frequencies, where the NOP is actually in the NDC region.
Recently, some of us showed\cite{WingePRA2018}, that ignition occurs in the state of oscillating field domains and that the arising lasing field afterwards stabilizes the behavior, see Fig.~\ref{fig:average}(a). In the transition region, where lasing starts and coexists with domain formation, see Fig.~\ref{fig:average}(c), our numerical simulations provide interesting complex dynamics including chaos, which we analyse in detail here. While chaotic behavior had been recently found in QCLs under external periodic driving \cite{JumpertzLight2016,SpitzScientificReports2019}, we note, that our system is autonomous.

Our article is organised as follows: In Sec.~\ref{sec:Model} we briefly repeat our model detailed in \cite{WingePRA2018}.
Here we focus on the differential equations describing the time evolution of electric fields $F_k$ in the modules and the occupations of the relevant lasing modes $N_\textrm{ph}^i$, which are our main variables. Section~\ref{sec:Inhom} presents a detailed analysis of the spatio-temporal evolution in the QCL just after ignition. Subsequently, we show in Sec.~\ref{sec:Lyapunov}, that the irregular behavior observed exhibits positive Lyapunov exponents, which proves that we observe chaos.

\section{\label{sec:Model}Model}

\begin{figure}
\centering
\includegraphics[scale=0.6]{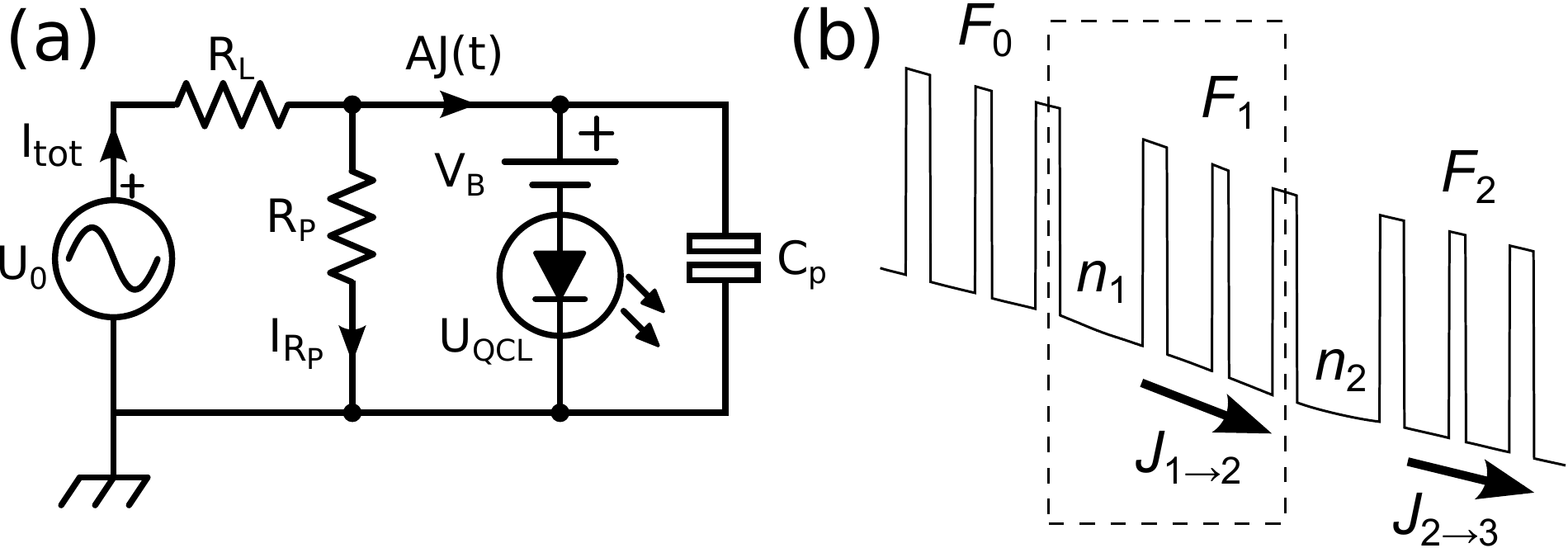}
\caption{(a) Electrical circuit design including probe resistance $\mathrm{R_p}$ and load resistance $\mathrm{R_L}$, Schottky potential $\mathrm{V_B}$ and parasitic capacitance $\mathrm{C_p}$ parallel to QCL. (b) Illustrated view of $3$ well design of QCL with current densities $\mathrm{J_m}$, electron densities $\mathrm{n_m}$ and fields $\mathrm{F_m}$ with module number $\mathrm{m}$ respectively. (Figure is modified from Ref.~\citep{WingePRA2018}.)
\label{fig:circuit}}
\end{figure}

In order to study the formation of field domain formation, we consider the dynamical evolution of the (average) electric field $F_m$ in module $m$ of the QCL containing $N=222$ modules of thickness $d=45.07$ nm, see  Fig.~\ref{fig:circuit}(b). In full analogy to superlattices \cite{WackerPhysRep2002,BonillaRepProgPhys2005} and earlier QCL studies \cite{WienoldJAP2011}, we have
\begin{equation}\begin{split}
\epsilon_r\epsilon_0\frac{dF_m}{dt}=&J(t)-J_{m\to m+1} \\
& -\frac{C_p}{C_s+C_p}\left(J(t)-\frac{1}{N+1}\sum_{k=0}^NJ_{k\to k+1}\right)
\end{split}
\label{eq:efield}
\end{equation}
where $\epsilon_r=12.9$ is the average relative permittivity, $C_s$ is the capacitance of the QCL structure and $C_p$ is a parasitic capacitance in parallel to the device, see Ref.~\citep{WackerJAP1995} for a derivation. In the following we assume $C_p=C_s/4$. The current $J(t)$ is fed via the circuit shown in Fig.~\ref{fig:circuit}(a) resulting in
\begin{equation}
AJ(t)=\frac{U_0-U_{QCL}(t)-V_B}{R_L}-\frac{U_{QCL}(t)+V_B}{R_p}
\label{eq:current}
\end{equation}
where
\begin{equation}
U_{QCL}(t)=\sum_{m=0}^{N}F_m(t)d 
\label{eq:uqcl}
\end{equation}
is the total bias drop over the QCL (including a boundary region with field  $F_0$). Here, $U_0$ is the external bias applied to the device, which is the control parameter for our system. From the experimental setup\cite{WingePRA2018} we extract a load resistance $\mathrm{R_L}=41.2\,\mathrm{\Omega}$ and a probe resistance $\mathrm{R_p}=1050.4\mathrm{\Omega}$. Finally, $A=0.15\,\mathrm{mm}^2$ is the cross section of the QCL and $V_B=0.8$ V reflects the bias drop due to a Schottky barrier at the metal-semiconductor contact.
 
The currents $J_{k\to k+1}$ between the modules are determined as
follows: For $k=1\ldots N-1$, we use an expression $J_{k\to
  k+1}(F_k,n_k,n_{k+1},\{N_\textrm{ph}^i\})$ based on our
non-equilibrium Green's function (NEGF) scheme
\cite{WackerIEEEJSelTopQuant2013} (using $7$ states per module). The
homogeneous results in Fig.~\ref{fig:average}(a) show $J_{k\to
  k+1}(F_k,n_D,n_D,\{N_\textrm{ph}^i\})$ for the  areal doping density
of $n_D=3\times 10^{10}/\mathrm{cm}^2$ (located in the center of the
largest wells). For the inhomogeneous case, the areal electron density
in module $k$ is given by
\begin{equation}
n_k=n_{D}+\frac{\epsilon_{r}\epsilon_{0}}{e}(F_{k}-F_{k-1})\, .
\end{equation}
Details are given in Ref.~\citep{WingePRA2018}. The boundary currents
at the beginning and the end of QCL structure are estimated by a
phenomenological conductivity $\sigma=0.15$ $\mathrm{A/Vcm}$ using
$J_{0\to 1}=\sigma F_0$ and $J_{N\to N+1}=n_N\sigma F_N/n_D$ in
analogy to Refs.~\citep{WackerPhysRep2002,WienoldJAP2011}.  We use a
lattice temperature of 77~K throughout this work. 

The occupations $N_\textrm{ph}^i$ of the cavity electromagnetic modes
$i$ with frequency $\omega_0^i$ are changing due to the interplay by
gain $G$ from the QCL medium and waveguide/mirror losses (quantified
by the threshold gain $g_{th}=20/\mathrm{cm}$) as
\begin{equation}
\frac{dN_\mathrm{ph}^{i}(t)}{dt}=\left(G(\omega^i_0)-g_{th}\right)\frac{c}{n_{g}}N_{ph}^{i}(t) +\sum_{k}\frac{An_{k}^{ULS}}{\tau_{sp}^{i}}
\, , 
\label{eq:photon}
\end{equation}
where $n_g=3.6$ is the group refractive index assumed to be constant
here. We also considered spontaneous emission with a time
$\tau_{sp}^{i}=3$ ms, where $n_{k}^{ULS}$ is the areal carrier density
in the upper laser level in module $k$. The gain $G(\omega)$  is the
sum of the gain contributions $G(F_k,\omega,\{N_{ph}^{i}\})$ for all
modules $k$, which are extracted from our NEGF calculations, see
Ref.~\citep{WingePRA2018} for details. 

Our model provides a closed system of equations for the fields $F_k$
with $k=0,1,\ldots N$ and the photon occupations  $N_\mathrm{ph}^{i}$,
where we have $38$ relevant modes with frequencies between
$10$ and $16$ $\mathrm{meV}$ in the cavity. We tacitly assumed that
the internal electron dynamics inside the QCL is instantaneously
adapting to the actual fields and mode occupations. This is probably a
good approximation, as typical scattering times are shorter than 1
ps. In comparison, the photon lifetime is $n_g/(g_{th}c)=6$ ps and the
dielectric relaxation time is at least $\epsilon_r\epsilon_0 (d F/d J)
= 5$ ps, based on the maximal slopes in Fig.~\ref{fig:average}(a). 

Fig.~\ref{fig:average}(c) shows time averaged results for bias and current upon simulating Eqs.~(\ref{eq:efield}-\ref{eq:photon}) for different values of our control parameter $U_0$. These results
(essentially identical with data shown Ref.~\citep{WingePRA2018} except for a slight improvement in the numerics) agree well with experimental measurements as detailed in Ref.~\citep{WingePRA2018}.
This shows the validity of our simulations, which have no fit parameters except for the contact conductivity and assuming a higher lattice temperature than in the experiment, which mimics heating of the phonon distribution \cite{VitielloAPL2012,ShiJAP2014}.

\begin{figure*}
    \centering
    \includegraphics[scale=0.35]{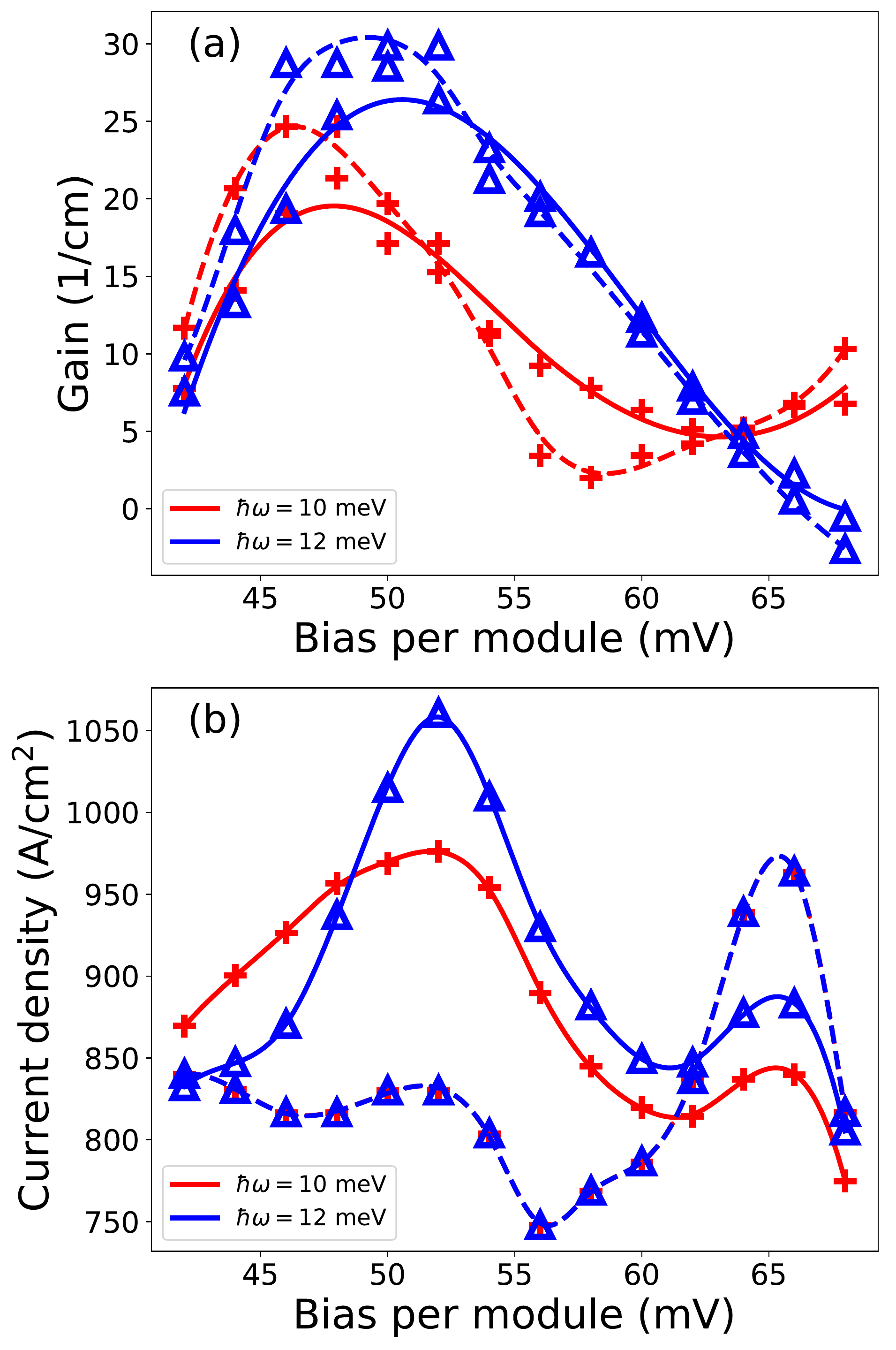}
    \hfill
    \includegraphics[scale=0.35]{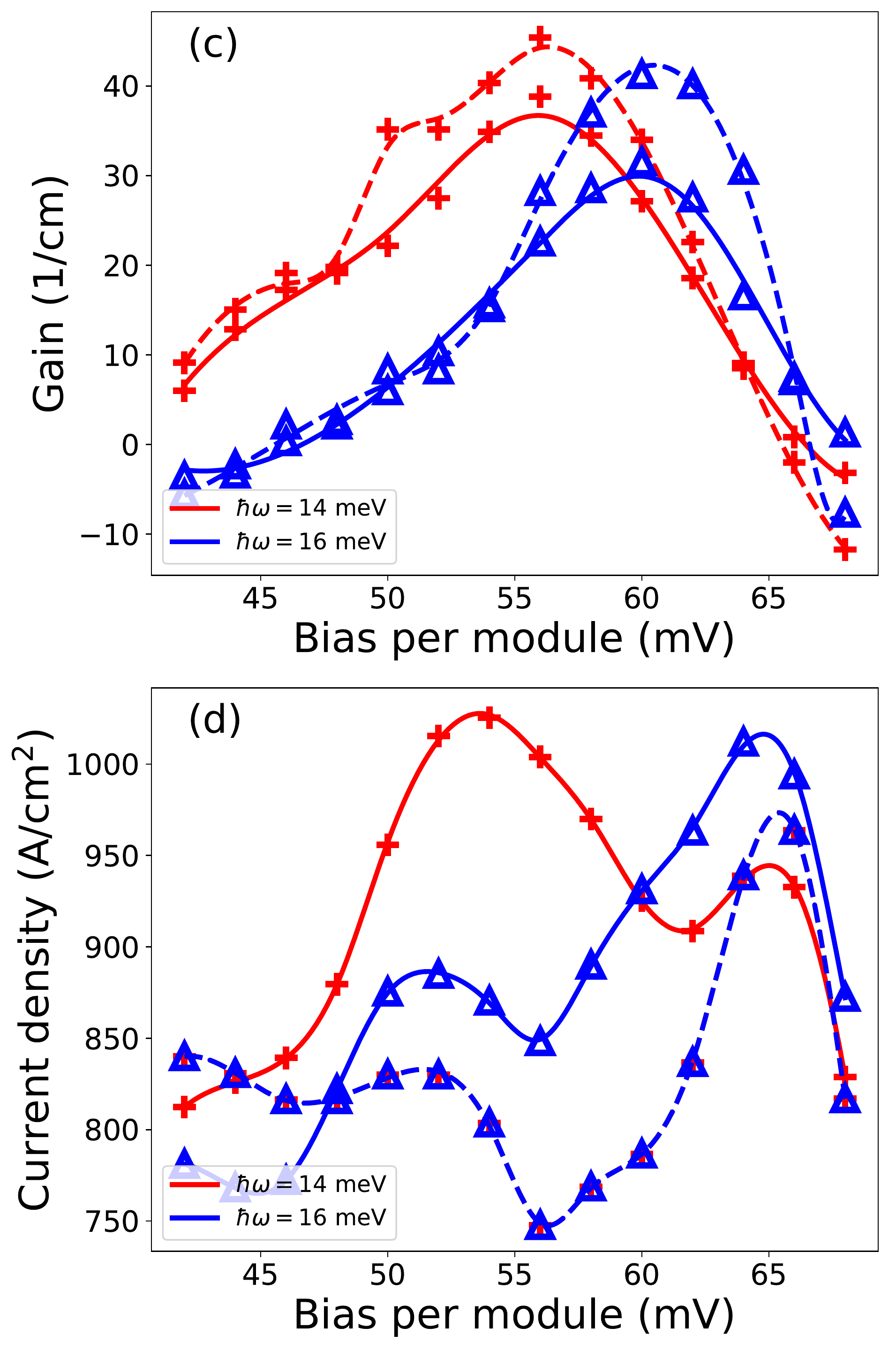}
    \caption{Gain (a,c) and current density (b,d) for different operation conditions extracted from our NEGF program. Simulations are done with two intensities: linear response at $eF_\mathrm{ac}d=0.1$ $\mathrm{meV}$ and nonlinear response at $eF_\mathrm{ac}d=10$ $\mathrm{meV}$ where the ac field strength reflects the photon number in the cavity. These are connected by dashed and solid lines, respectively, to guide the eye.  In each panel, results for low(high) frequencies with photon energies in the range between $\hbar\omega=10$ $\mathrm{meV}$ and $\hbar\omega=16$ $\mathrm{meV}$ are shown as crosses(triangles).} 
    \label{fig:gainfit}
\end{figure*}

The central input to the model are the functions for gain and current, which show a wide variation with the system parameters as displayed in  Fig.~\ref{fig:gainfit}.
 The detailed fitting process for the gain and the current is discussed in \citep{WingePRA2018}.
 The mode frequencies chosen for these plots span the whole range and demonstrates that the data are highly frequency-dependent. For the linear response, the current density is largely unchanged. For higher intensity, gain saturation is observed together with an increase in the current due to the stimulated intersubband transitions.

\section{\label{sec:Inhom} Oscillating Field Domains and Chaos }

\begin{figure}
\centering
\includegraphics[scale=0.5]{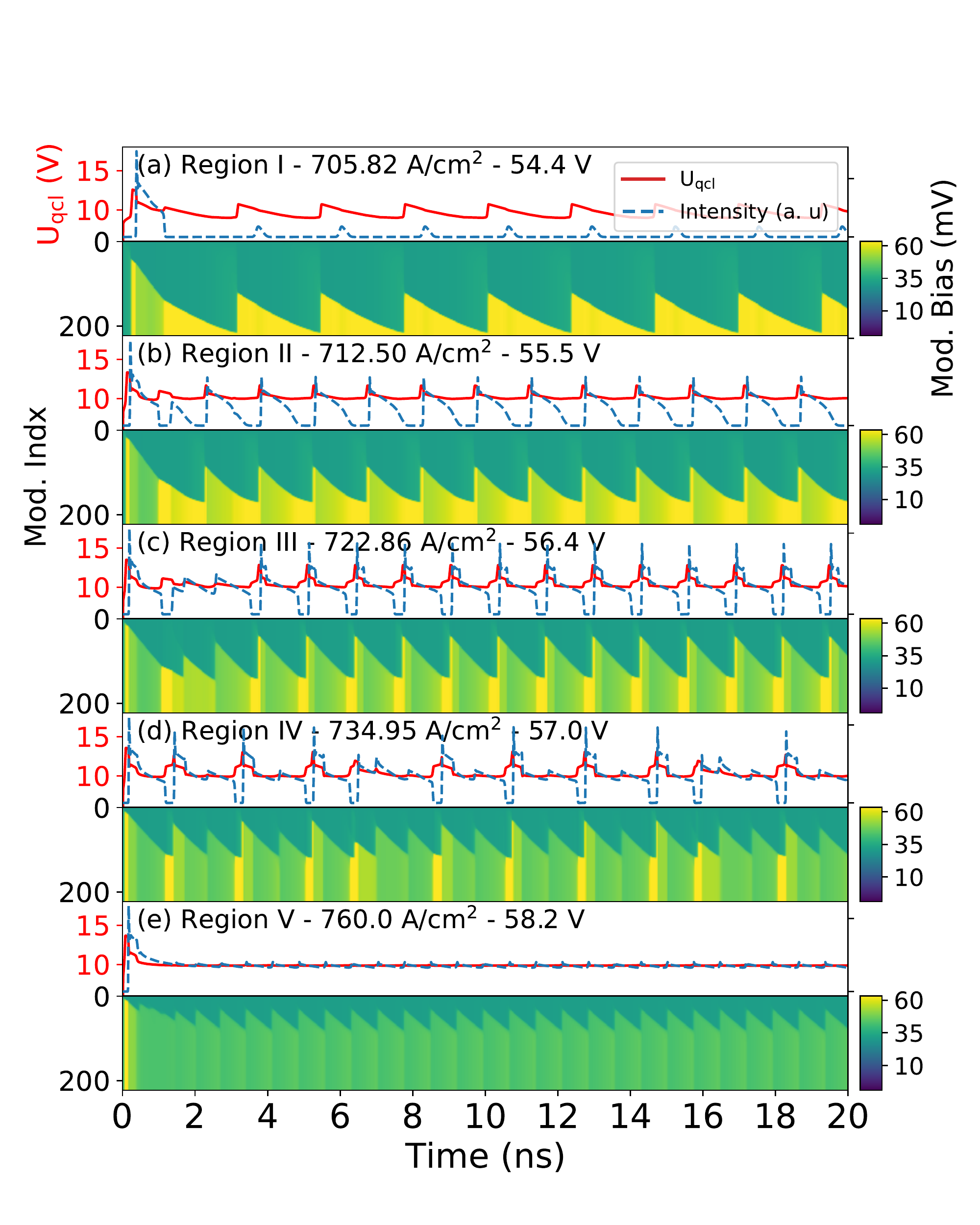}
\caption{\label{fig:field} Dynamic evolution for the initial 20 ns for different external biases $U_0$ as specified at the end of the upper line for each panel (a)-(e). The color-scale plots show the local bias drop in each module (vertical axis) as a function of time (horizontal axis). Above these, the QCL bias is shown by a full red curve and intensity by a dashed blue line.}
\label{fig:fields}
\end{figure}

In this section, we analyse the dynamical behaviour in the region shown
in Fig.~\ref{fig:average}(c) in detail. When $U_0$ becomes larger than $54.2$ V, the operation point reaches the NDC region for a homogeneous field distribution. This causes the formation of field domains with boundaries travelling through the device. This provides oscillations in current and bias and therefore the corresponding time-averages have been plotted in Fig.~\ref{fig:average}(c).
Without lasing, Fig.~\ref{fig:average}(a) shows that the condition of equal current density of about 700 A/cm$^2$ provides a bias drop per module of about 63 mV in the high-field domain and 36 mV in the low-field domain. From
Fig.~\ref{fig:gainfit}(c) we obtain a substantial gain at $\hbar \omega=16$ meV in the high-field domain, while there is only little absorption or gain for bias drops corresponding to the low-field domain. The amplification of the optical field can surpass the losses, if the high-field domain extends over a major part of the device. Thus, lasing sets on in a state of oscillating domains, see the dashed line in Fig.~
\ref{fig:average}(c). This lasing field strongly modifies the current and gain as demonstrated in Fig.~\ref{fig:gainfit} which results in the complex behaviour we observed. Here, we identified 5 distinct regions with essentially different behaviour as shown in Fig.~\ref{fig:fields}. 

 In region I, Fig.~\ref{fig:fields}(a) shows the characteristic oscillations due to travelling field domain boundaries.  The homogeneous field distribution becomes unstable (e.g. at $t=12$ ns) and splits up in a high- and low field domain. Afterwards the electron accumulation layer separating both domains travels to the positive contact and the scenario repeats, after the field in the low-field region has increased to maintain the bias. Just after its formation, the high-field domain can be large enough to provide sufficient gain to compensate losses, but this holds only for a short time, so that the lasing intensities are never sufficient to effect the behaviour in this region.

With increasing bias $U_0$ the high-field domain becomes more extended and the lasing field becomes stronger, so that it significantly enhances the current around 63 mV, see Fig.~\ref{fig:gainfit}(d). In order to keep the current density, the field in the high-field domain needs to diminish as can be seen Fig.~\ref{fig:fields}(b), which is characteristic for region II, where the average current increases stronger with bias than in region I. However, with shrinking length of the  high-field domain, the gain drops and the original bias per module of 63 mV is restored in the high-field domain before a new instability appears in the low-field domain associated with a peak in current.

 In region III, as the external drive increases, the lasing intensity becomes more strong and covers a wider range of each domain cycle, as seen in Fig.~\ref{fig:fields}(c). However, lasing always stops before a new domain boundary forms. In the local $U_{qcl}-J$ relation (see Fig.~\ref{fig:average}(c)), $U_{qcl}$ stabilizes while the intensity continues increasing.

In region IV, lasing starts to persist most of the time and some high field domains form around the NOP and the average bias drops as seen in Fig.~\ref{fig:fields}(d). Finally, in region V, lasing persists all the time and all the high field domains form around the NOP. $U_{qcl}$ almost stabilizes as the current increases slightly with the intensity. (see Fig.~\ref{fig:fields}(e) and Fig.~\ref{fig:average}(c)).

\begin{figure}
\centering
\includegraphics[scale=0.4]{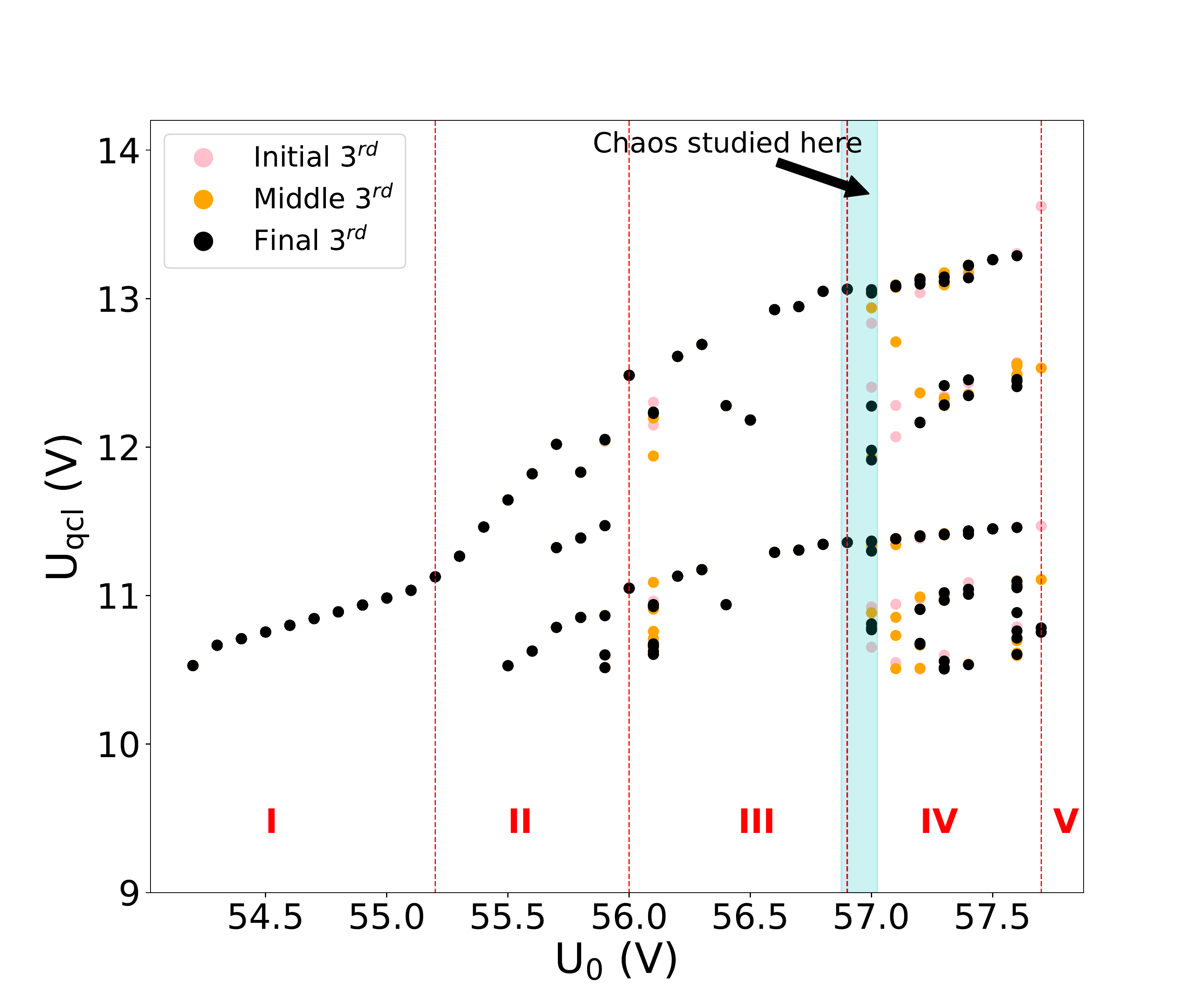}
\caption{\label{fig:bifur} Local maximas of the QCL voltage output with applied external bias where the dots represent the peaks of the time series (see Fig.~\ref{fig:time}). Three different color indicates three different parts of the data.
}
\label{fig:bifurcation}
\end{figure}

 As discussed in Fig.~\ref{fig:fields}, the system shows fundamentally different behavior in the five regions. In order to check for periodicity, we plot the local maxima of the $U_\textrm{qcl}(t)$ signal as a function of the control parameter in Fig.~\ref{fig:bifurcation}. Such a diagram is well-known for identifying routes to chaos \cite{JamitzkyNanotech2006,JumpertzLight2016}.
In order to identify irregular behavior, the time series $U_{qcl}(t)$ is divided into three equal interval in times marked with colors pink, orange and black after removing initial behavior in the first $160$ $\mathrm{ns}$. The black dots are from the latest time interval and cover all earlier dots with the same value. The orange dots are from the middle interval and their persistence indicates non-periodic solutions.  Finally, pink dots from the first period in time indicate that periodicity had not been reached within the first $160$ $\mathrm{ns}$, and their presence without yellow dots indicates long transients rather than chaos
\citep{TelChaos2015}. 

 \begin{figure}
\centering
\includegraphics[scale=0.4]{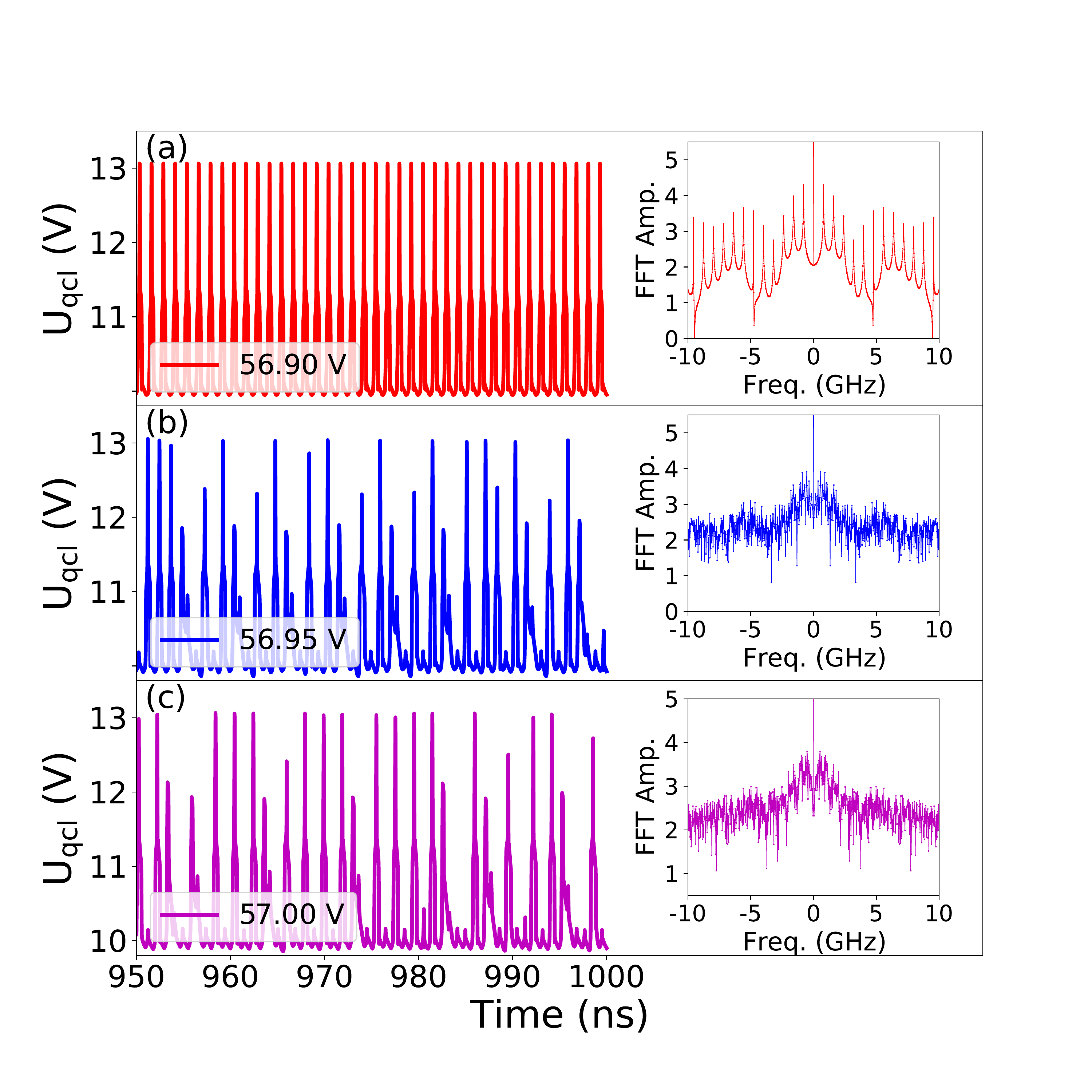}
\caption{ Time series of three $U_{QCL}$ data following their transition from regular to irregular region are shown in panel (a), (b) and (c) (blue highlighted region in Fig.~\ref{fig:bifurcation}). FFT of the time series are shown in the right inset. Oscillations observed in GHz range.}
\label{fig:time}
\end{figure}

 In region I, consistently with Fig.~\ref{fig:fields}(a), oscillations are regular. A couple of points in region II exhibit regular period two and three behaviour. In region III, there is a point $U_0$ at $56.1\mathrm{ V}$ that shows some irregularity which we did not analyse further. In general $U_{qcl}$ solutions continue to be regular. Shortly after the transition between region III to IV, we have strong indications of chaos among the region IV. The region V again has a regular structure. In the following we focus on the transition range which is highlighted as cyan color background. 

 In Fig.~\ref{fig:time}, we show the $U_{qcl}$ time series for three $U_0$ points from the cyan highlighted region in Fig.~\ref{fig:bifurcation}. These three $U_0$ points are chosen to investigate the transition from region III to region IV in more detail. 
For $U_0=56.9$ V be observe clear periodic behaviour with two peaks in Fig.~\ref{fig:time}(a). In contrast irregular behaviour is observed for slightly larger biases $U_0=56.95$ V and $U_0=57$ V as shown in panels (b) and (c). These irregular patterns are the first indications of chaos \citep{SchusterBook1988}.


 In Fig.~\ref{fig:time}, Fast Fourier Transforms (FFT) of the corresponding three $U_{0}$ points are shown in the insets locate on the right of the panels with oscillations in GHz range. Here, in the transition between regular to irregular oscillations, the structure clearly becomes continuous rather than discrete. This is not a proof of chaos but an indication consistent with the time series.

\begin{figure}
\centering
\includegraphics[scale=0.4]{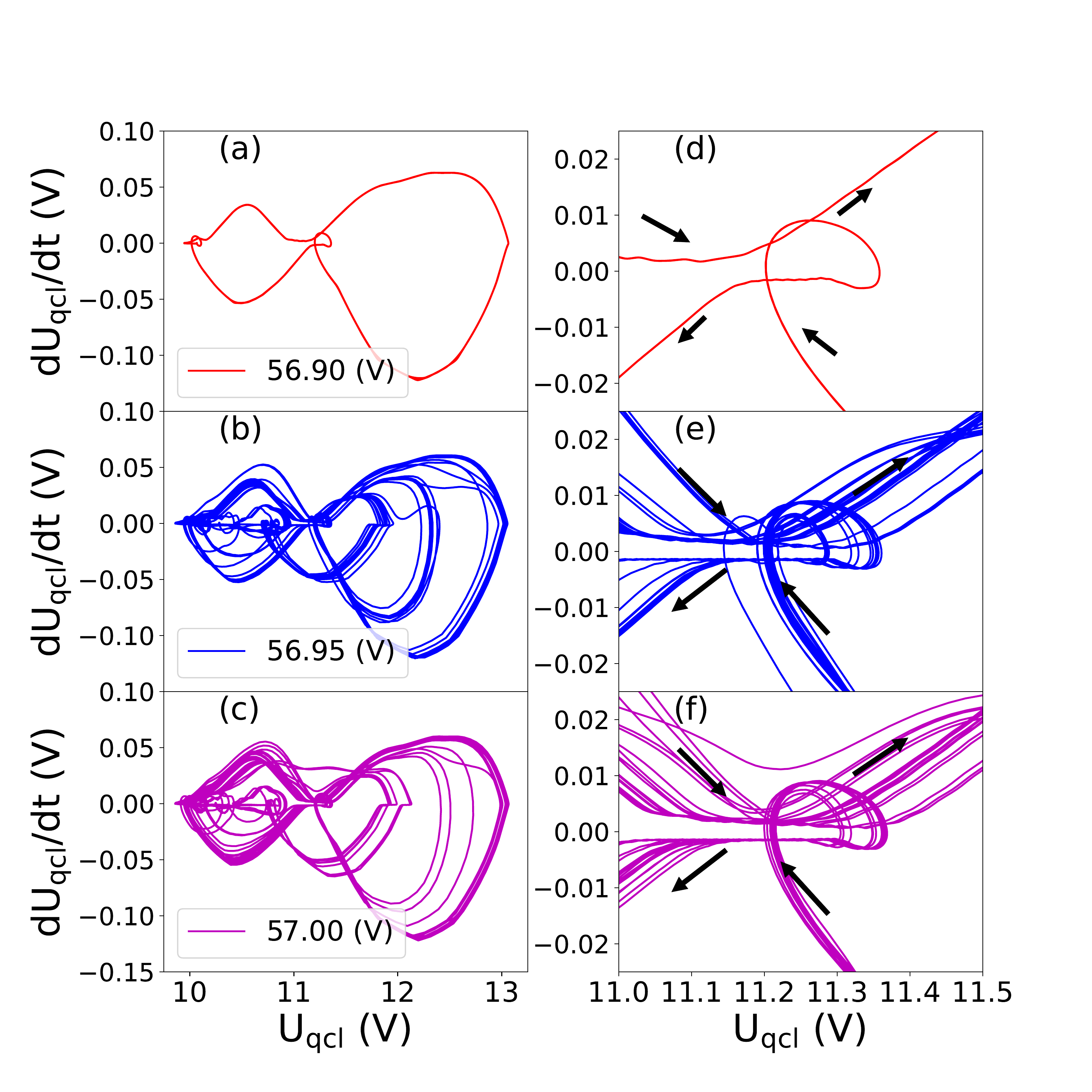}
\caption{ In panel (a), (b) and (c) phase diagrams of the time series for different bias are shown. In panels (e), (f) and (d) the phase diagrams are enlarged around tangential points to show the separation of the flows for different paths. Arrows represent the flow direction.}
\label{fig:phase}
\end{figure}

 With constructing the time series and FFT of the three $U_{qcl}$ data discussed in Fig.~\ref{fig:time}, we will show these rigorous numerical results providing chaos.  
The most common methods of proving chaos are usually constructing the phase spaces and deriving the Lyapunov exponents. To understand the dynamical evolution of the system, phase diagram is one of the main steps to investigate the chaos \citep{GolubitskyBook2002}.

 To analyze the results we have chosen $U_{qcl}$ and the time derivative of $U_{qcl}$ as two variables out of our complex system with 223 variables to define a new reduced phase diagram. For simplicity, this is referred to as phase diagram in the following. In Fig.~\ref{fig:phase}, phase diagrams of the three $U_{0}$ data from Fig.~\ref{fig:time} are shown. In Fig.~\ref{fig:phase}(a) and (d), one can clearly see that the regular data is following the same path in each period without any deviation. Therefore, small perturbations of the data do not change any dynamics of the system. However, In Fig.~\ref{fig:phase}(b), (c), (e) and (f) one can clearly observe the deviations of the path spread around the phase diagram. While the trajectories come close to each other at some places, tiny differences grow to qualitative different behaviour in the course of the time-evolution, as characteristic for chaos.  There are some similar phase diagram constructions done in other works including tangential junctions with stable and unstable manifolds (see Ref.~\cite{GavrilovMathUSSRSB1972,GavrilovMathUSSRSB1973,WilliamsonChaos2015,BertozziSiamJMathAnal1988,BorisovChaoticDyn2016}).

\begin{figure*}
    \centering
    \begin{minipage}{0.41\textwidth}
        \centering
        \includegraphics[scale=0.5]{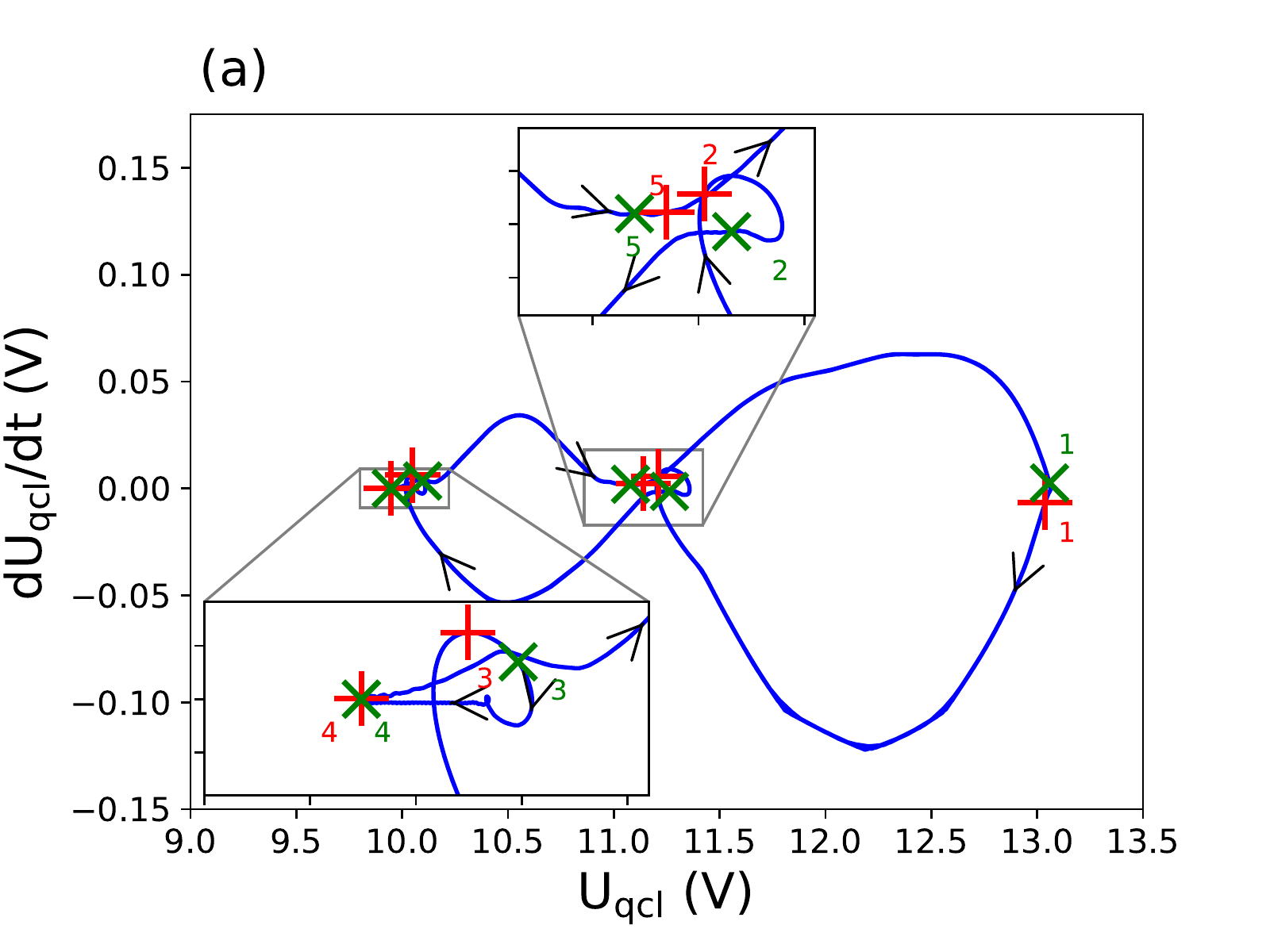}
    \end{minipage}\hfill
    \begin{minipage}{0.54\textwidth}
        \centering
        \includegraphics[scale=0.35]{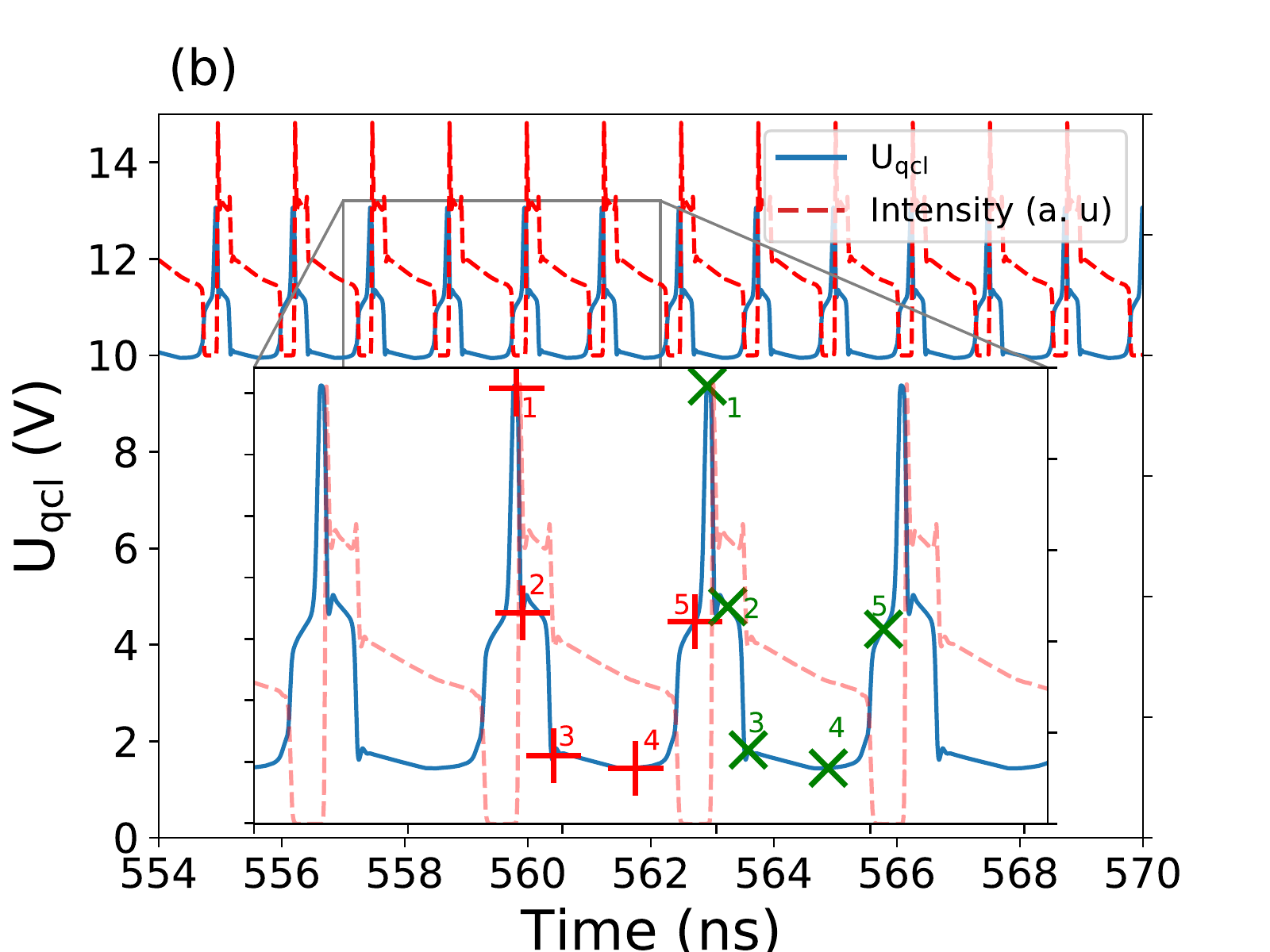}
        \includegraphics[scale=0.35]{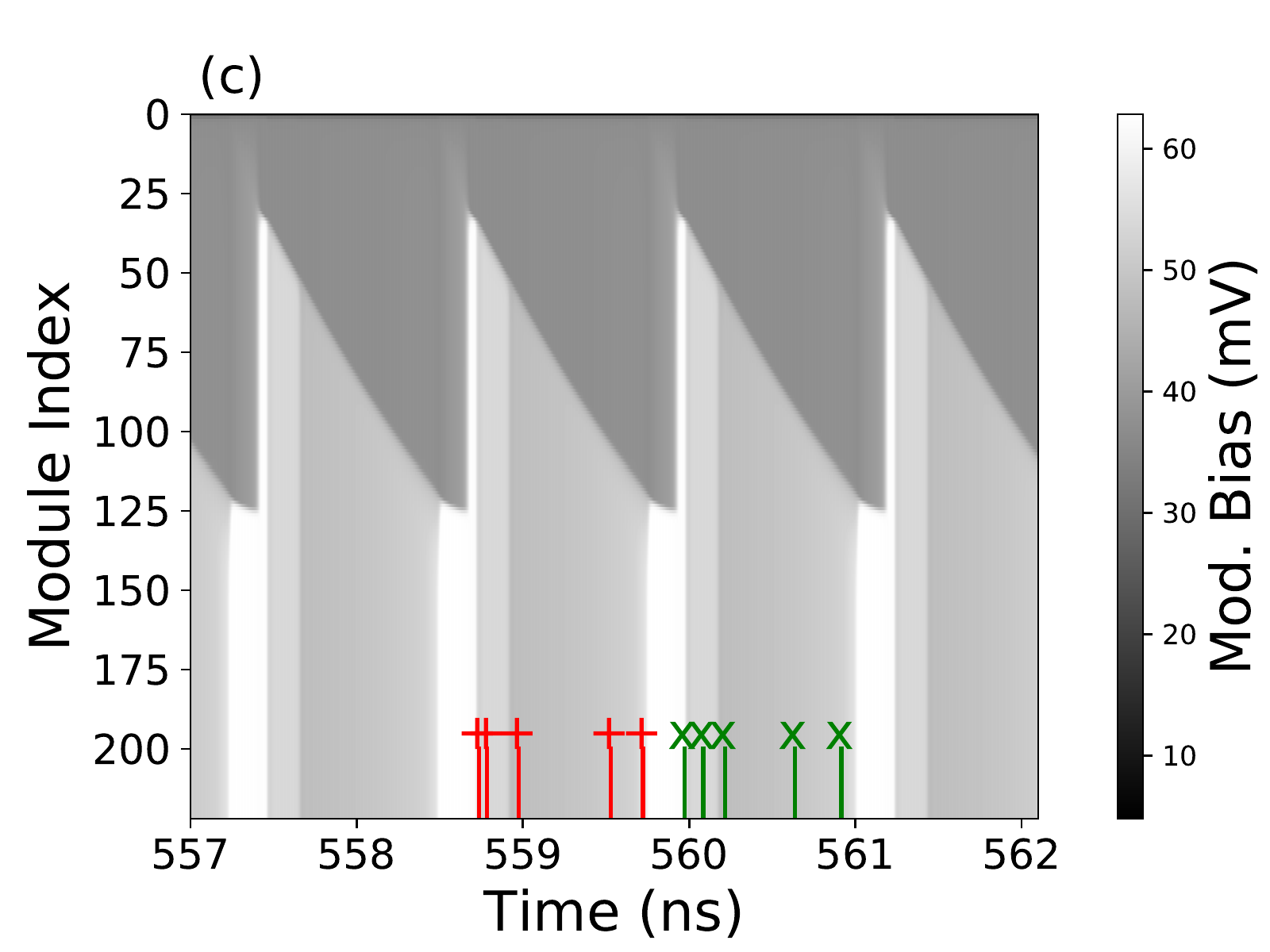} 
    \end{minipage}
    \caption{Phase diagram, time series and domain oscillations for $U_0=56.90$ $\mathrm{V}$ . In panel (a) the trajectory  is shown for the time-window in panel (c) as a highlighted blue line. Color markers allow the comparison of specific times with panels (b) and (c).  In panel (b) time series of bias (blue curve) and intensities (dashed red curve) are shown. The enlarged range is marked consistently with panel (a) and (c). In panel (c) the field distribution among the modules is shown. }
    \label{fig:phase5690}
\end{figure*}

In Fig.~\ref{fig:phase5690} we analyse the behavior for the periodic oscillation at $U_0=56.90\mathrm{ V}$ in detail. The phase diagram in panel (a) shows, that two subsequent periods (marked by red plus and green cross symbols)
  lie on top of each other. The lasing intensity [red dashed line in panel (b)] is essentially dropping to zero around symbol 5  before the new domain boundary forms, which is associated with a sharp peak in the $U_\textrm{qcl}(t)$ signal at symbol 1. A smaller bias peak arises close to symbol 2 just after the domain formation. This behavior is typical for region III as discussed in Fig.~\ref{fig:average}(c) and Fig.~\ref{fig:bifurcation}.  

\begin{figure*}
    \centering
    \begin{minipage}{0.42\textwidth}
        \centering
        \vspace{0cm}
        \includegraphics[scale=0.5]{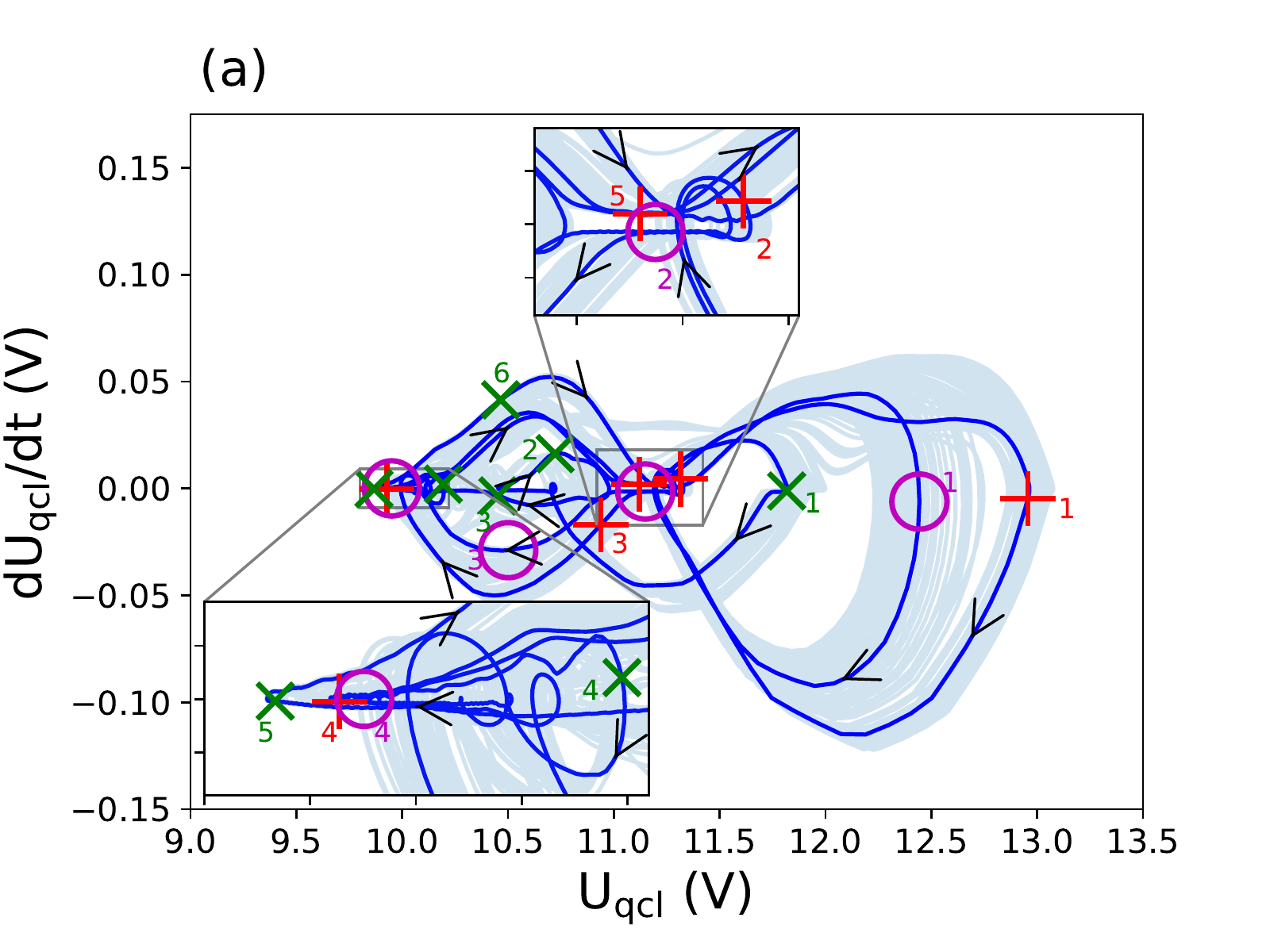}

        
    \end{minipage}\hfill
    \begin{minipage}{0.54\textwidth}
        \centering
         \includegraphics[scale=0.35]{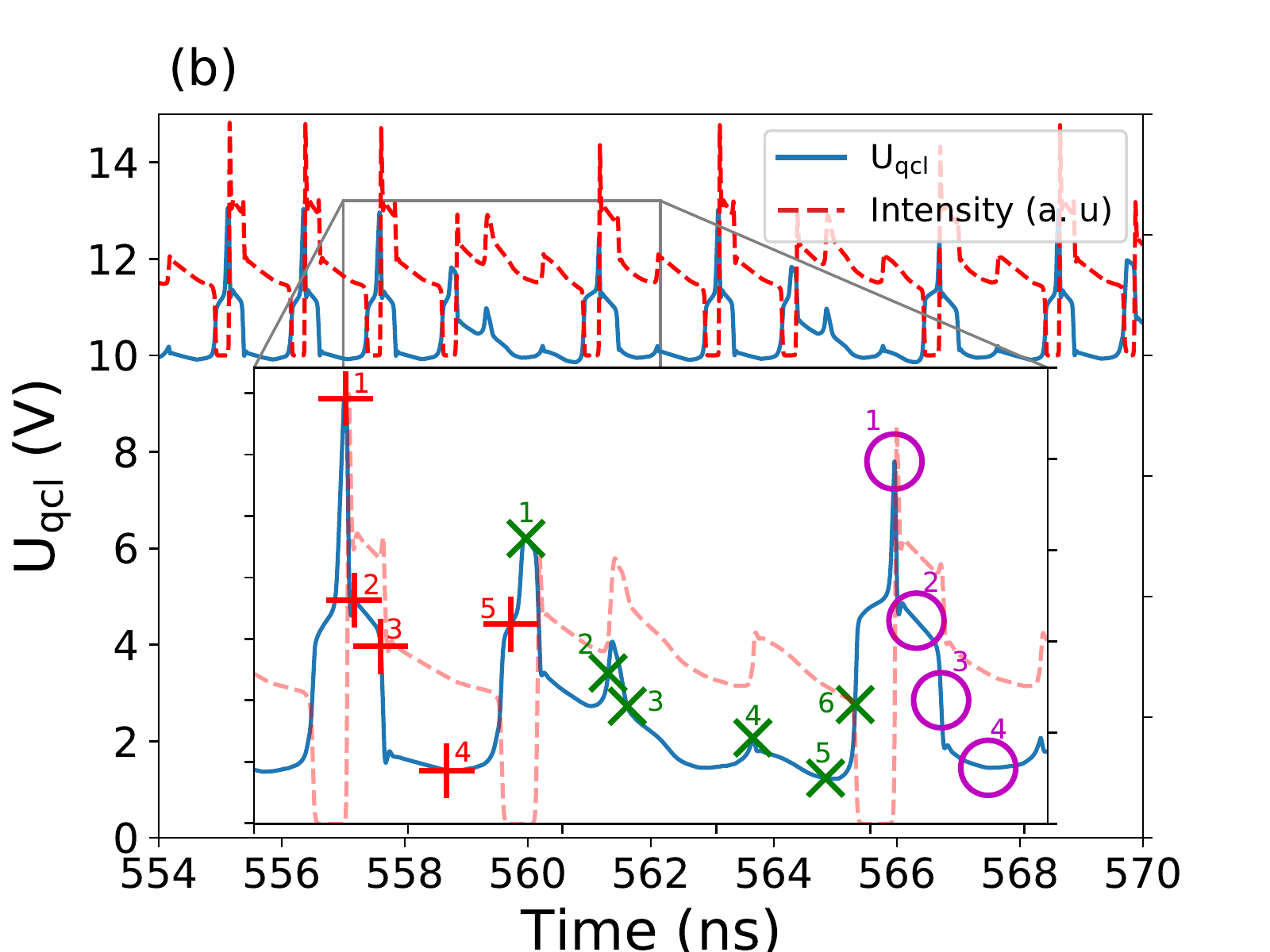}
         \includegraphics[scale=0.35]{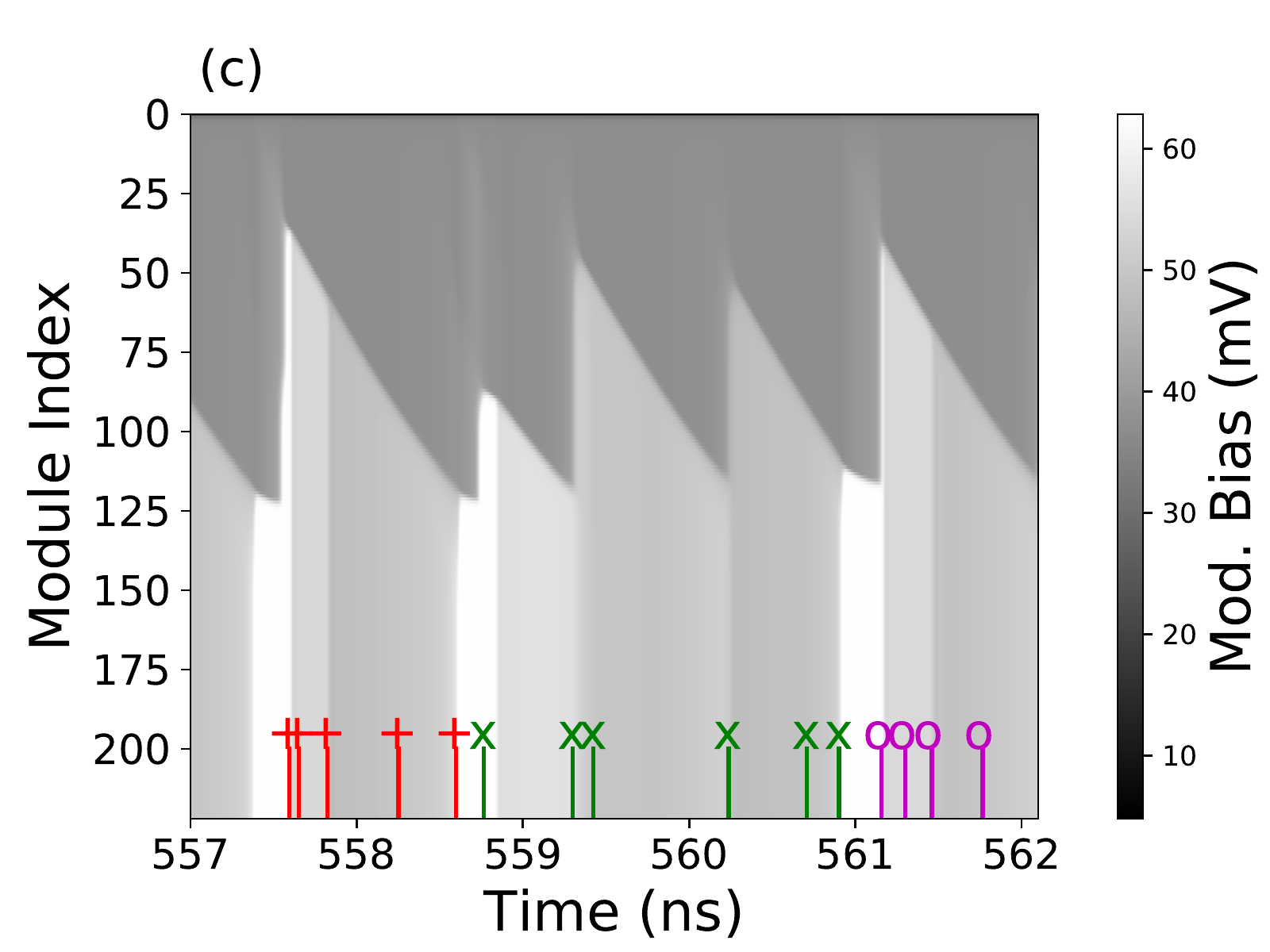} 
    \end{minipage}
    \caption{Phase diagram, time series and domain oscillations for $U_0=56.95$ $\mathrm{V}$. In panel (a) the trajectory  is shown for the time-window in panel (c) as a highlighted blue line.  Lower opacity lines on the background represent a longer time series. Color markers allow the comparison of specific times with panels (b) and (c).  In panel (b) time series of bias (blue curve) and intensities (dashed red curve) are shown. The enlarged range is marked consistently with panel (a) and (c). In panel (c) the field distribution among the modules is shown. }
    \label{fig:phase5695}
\end{figure*}

As the external drive increases into the region IV, more complicated and irregular features start to appear as seen in Fig.~\ref{fig:phase5695}. Here the same method is used to plot the variables as in Fig.~\ref{fig:phase5690} but with three major peaks of bias marked with 1. Three markers (red plus, green cross and magenta circle) now represent the trajectories after the respective bias peak in the phase diagram . We see, that the trajectories deviate significantly in each cycle. These deviations bring some new scenarios in the trajectories. While the red and magenta markers follow a similar path with a significant deviation, the green trajectory stays on a whole different path. Eventually all the markers get together around $9.75\mathrm{ V}$ in Fig.~\ref{fig:phase5695}(a). It appears this point of junction is the place where the system decides how to evolve. Also, as seen in Fig.~\ref{fig:phase5695}(b) and (c) lasing stays persistent even while
  new domain boundaries form in the middle part with green cross markers. In contrast, at the major bias peaks,
  the domain boundaries form a state with vanishing intensity.

\section{Lyapunov Exponents \label{sec:Lyapunov}}
 As we seek to quantify the presence of chaos in the QCL we consider the largest exponent from the Lyapunov spectrum of the system \cite{ShawZNA1981,StrogatzBook2015}. Let $d(t)$ be the distance between two closely lying state vectors in a bounded phase space at time $t$ then the largest Lyapunov exponent $\lambda$ describes the time-evolution of the distance
\begin{equation}
    d(t) \sim \mathrm{e}^{\lambda t}d(0)
\end{equation}
When $\lambda$ is positive, small deviations in the state vector lead to exponentially growing deviations in phase space, i.e. chaos (provided the phase space is bounded).

To estimate $\lambda$ from the time series we employ a procedure similar to the ones in Refs. \cite{SatoProgThPhys1987,RosensteinPHD1993} albeit with the difference that we have access to the true phase space of the system, i.e. the state of the modules, hence, we can ignore the question of how to reconstruct the attractor.
Consider the field of the $m$'th module with time series $F_m(t)$  and define its normalized variable 
\begin{equation}
    v_m(t) = \frac{F_m(t)-\langle F_m\rangle}{\mathrm{std}(F_m)}
\end{equation}
with $\langle F_m\rangle$ and $\mathrm{std}(F_m)$ being, respectively, the mean and standard deviation of $F_m(t)$ over time.
This yields the normalized state vector at time $t$: $v(t) = (v_1(t),v_2(t),\cdots,v_N(t))$.
From the normalization of the components of $v(t)$ the variation of every variable is included equally in the analysis.

For two state vectors at times $t_i$ and $t_j$, respectively, we define the Euclidean distance after the evolution time $t_\mathrm{evo}$ as
\begin{equation}
    d_{ij}(t_\mathrm{evo}) = |v(t_i+t_\mathrm{evo}) - v(t_j+t_\mathrm{evo})|
\end{equation}
When estimating the Lyapunov exponents we traverse a fiduciary trajectory of $v(t)$.
At time $t_i$ we search the time series for another time $t_j$ for which $d_{ij}(0)$ is minimal.
 Here, we exclude any state vectors that are within the interval $(t_i-\Delta;t_i+\Delta)$ to avoid trivial pairing between the state vector and its negligibly shifted self.

In a system of limited phase space volume the distance between any two state vectors is bounded.
Hence, the deviation between the pairs will saturate and the picture of exponentially increasing distances between any two close pairs only holds for small initial deviations.
This is ensured by restricting the sampling to pairs with initial distances below a certain discriminator $d_{ij}(0) < D$.

From the identified pairs $t_i$ and $t_j$ we estimate the contribution to the Lyapunov exponent at $t_\mathrm{evo}$ as
\begin{equation}
    \lambda_{ij}(t_\mathrm{evo}) = \frac{1}{t_\mathrm{evo}}\log\frac{d_{ij}(t_\mathrm{evo})}{d_{ij}(0)}
    \label{eq:lyapunov}
\end{equation}
The average contribution of all pairs $(i,j)$ yields the estimated Lyapunov exponent $\hat{\lambda}(t_\mathrm{evo}) =\langle{\lambda_{ij}}(t_\mathrm{evo})\rangle$.

The procedure above neglects the phase orientation information between one pair and the next when determining the sequence of pairs which is otherwise included in the algorithm of Wolf \textit{et al.} \cite{WolfPHD1985}.
Yet, as we are only interested in the largest Lyapunov exponent such phase information is unnecessary \cite{RosensteinPHD1993}.

\begin{figure*}
    \centering
    \includegraphics[width=0.45\textwidth]{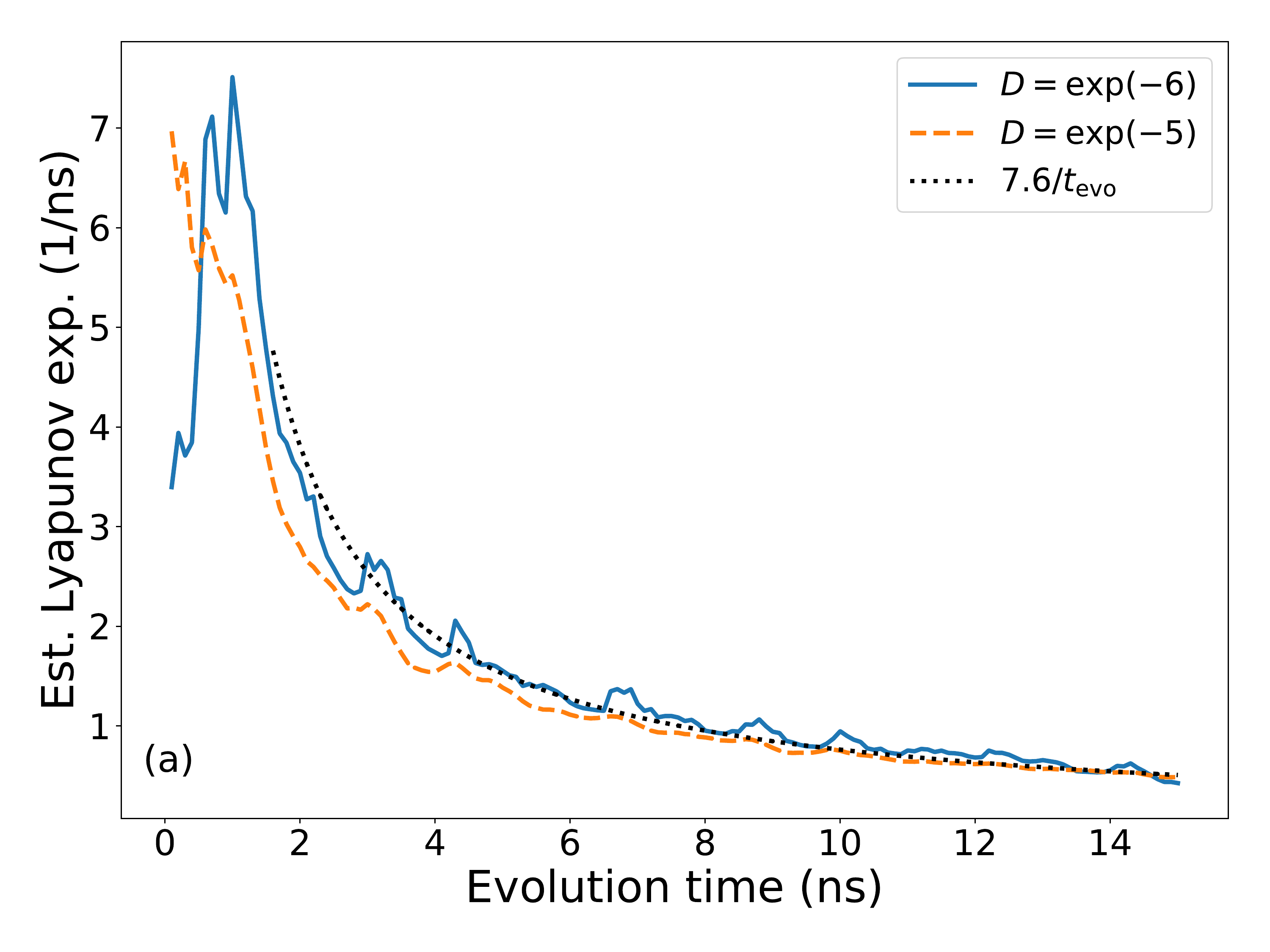}
    \hspace{2mm}
    \includegraphics[width=0.45\textwidth]{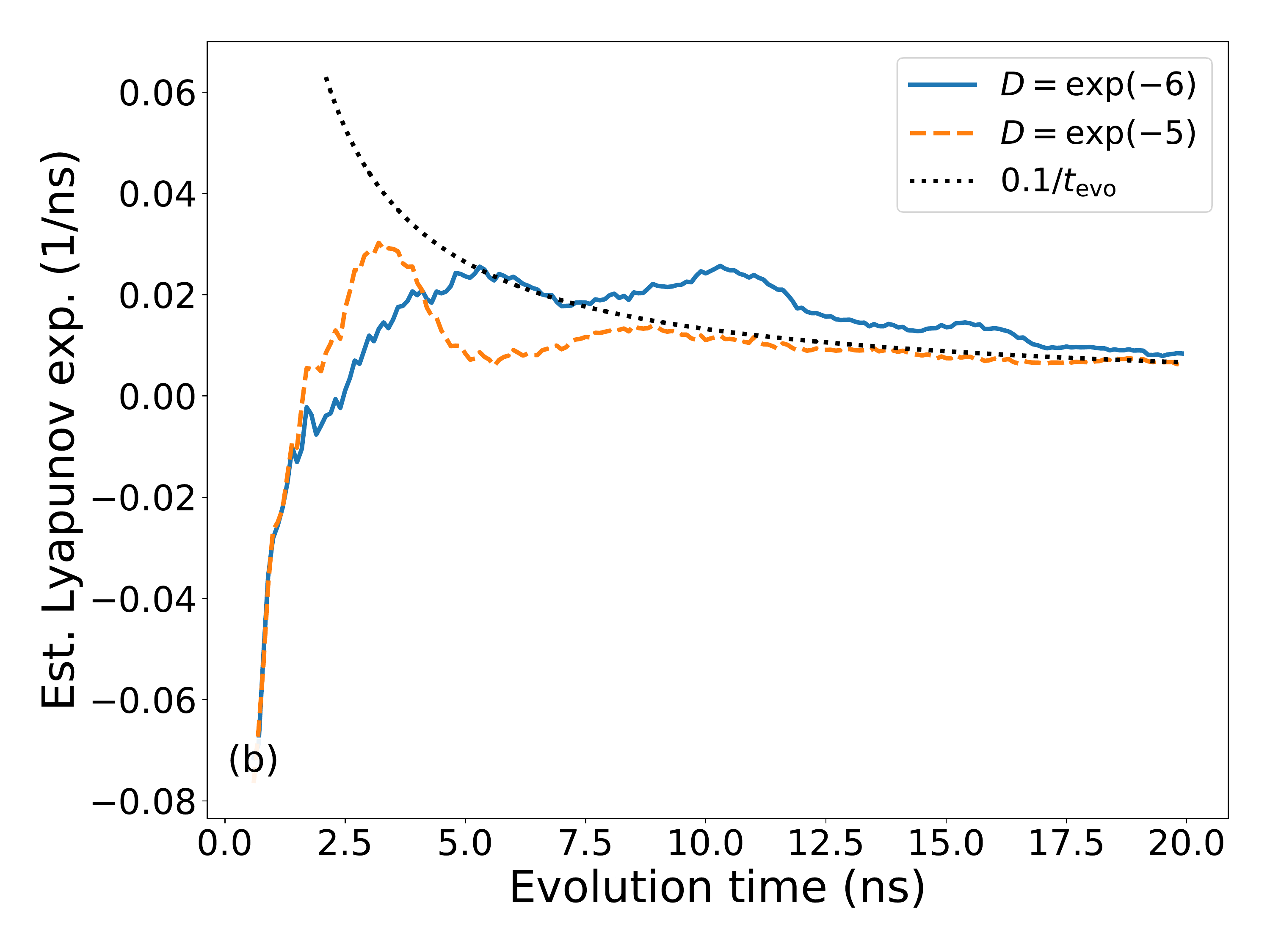}
    \caption{Lyapunov exponents for biases (a) 56.95 V and (b) 56.90 V. The exponents are estimated using the exclusion time $\Delta = 5$ ns and different discriminators $D$ for the solid blue and dashed orange lines, respectively. For larger evolution times the $1/t_\mathrm{evo}$ behaviour is shown by black dotted lines to guide the eye.}
    \label{fig:lyapunov}
\end{figure*}

Fig.~\ref{fig:lyapunov} provides the Lypaunov exponents for the biases 56.95 V and 56.90 V.
In both cases the exclusion time is $\Delta = 5\textrm{ ns}$ and we consider two different discriminators ${D = \exp(-6),\exp(-5)}$.
Considering first Fig.~\ref{fig:lyapunov}(a) with bias 56.95 V, the curves show that the estimated Lyapunov exponents (solid blue and dashed orange lines) exhibit a plateau for evolution times around 2 ns, the plateau resembles the stationary of the Lyapunov exponents for certain evolution times observed in in Refs. \cite{RouxPHD1983,WolfPHD1985,RosensteinPHD1993,SpitzScientificReports2019}.
The plateau yields a positive estimate for the Lyapunov exponent of the order of  $6\textrm{ ns}^{-1} > 0$ which corroborates our interpretation of the irregular pattern in Fig.~\ref{fig:phase5695} as chaotic behaviour.
Beyond the plateau the estimate decreases according to $1/t_\mathrm{evo}$-behaviour (the black dotted line), as expected from \eqref{eq:lyapunov} with saturation bound on the distance.  

Repeating the analysis for Fig.~\ref{fig:lyapunov}(b) with bias 56.90 V we find that the estimates for both discriminators are two orders of magnitude smaller than for Fig.~\ref{fig:lyapunov}(a) and hover around zero, in agreement with the regular behaviour observed in Fig.~\ref{fig:phase5690}.  

\section{\label{sec:conclusion}Conclusion and Outlook}
In this work, we analyzed chaotic behavior in a QCL without external  driving. This occurred for a particular QCL operating in its NDC region, where traveling field domains form. Here the high-field domain exhibited gain, resulting in pulses of light during periods, where this high-field domain is sufficiently large. With increasing driving, the pulses get more pronounced and can modify the subsequent formation of a new domain boundary. Here, two different formation scenarios exist and their succession becomes irregular, resulting in the chaotic signal. For even higher driving, the lasing never stops and we recover an ordinary laser operation.

In phase diagrams, we could identify the points in which the trajectories deviate from each other. Here we observe the sensitive dependence on initial conditions as demonstrated by positive Lyapunov exponents. Thus, QCLs form a further autonomous system of technological relevance showing chaos.

It would be interesting to study this experimentally for undriven QCLs. Indeed the device studied here showed subharmonics of period three\cite{WingePRA2018} at a single operation point, which can be seen as an indication for chaos \cite{LiAmMathMonthly1975}. More data are very welcome to reconstruct the phase space in similar devices which could unambiguously demonstrate chaos in these technologically important systems.
\section*{Acknowledgements}
We thank the Swedish Research Council (project
2017-04287) and NanoLund for financial support.


\end{document}